\documentclass{ws-ijmpa}

\newcommand\munit{$\mathrm{GeV}/c^2$}

\begin{document}

\markboth{P.~Wagner, D.~Toback}
{Prospects of a Search for Neutral, Long-Lived Particles using Photon Timing at CDF}

%
\catchline{}{}{}{}{}
%

\title{Prospects of a Search for Neutral, Long-Lived Particles using Photon Timing at CDF
}

\author{\footnotesize PETER~WAGNER\footnote{wagnp@fnal.gov, conference speaker}, DAVID~TOBACK\footnote{toback@fnal.gov}}

\address{Dept. of Physics, Texas A\&M University\\
College Station, Texas 77843-4242
}

\maketitle

\pub{Received (18 Oct 2004)}{}

\begin{abstract}
We present the prospects of searches for neutral, long-lived particles
which decay to photons using their time of arrival measured with a
newly installed EMTiming system at CDF. Using GMSB $\tilde{\chi}_{1}^{0}\rightarrow\gamma\tilde{G}$
models we estimate the expected
95\%~confidence level exclusion regions as a function
of the neutralino mass and lifetime. We find that a combination of
single photon and diphoton analyses should allow the Tevatron in run~II
to easily extend the exclusion regions from ALEPH at LEP~II, and cover parts of the theoretically
favored $m_{\tilde{G}} < \mathrm{few}$~$\mathrm{keV}/c^2$ GMSB parameter space.

\end{abstract}

~\linebreak

The electromagnetic (EM) calorimeter at CDF has recently been equipped with a new
timing system, EMTiming\cite{key-24}, to measure the arrival time
of energy deposited (e.g. from photons). While
it was initially designed to reject cosmics and accelerator backgrounds\cite{key-10}, 
we summarize the prospects of using it to search
for neutral particles with a lifetime of the order
of a nanosecond which decay in flight to photons\cite{key-ournote}. An example of
a theory which would produce these particles is gauge mediated supersymmetry
breaking (GMSB)\cite{key-2} with a neutralino, $\tilde{\chi}_{1}^{0}$, as the
next-to-lightest supersymmetric particle (NLSP) and a light gravitino,
$\tilde{G}$, as the LSP. In this scenario the neutralino
decays preferably ($\sim$100\%) as $\tilde{\chi}_{1}^{0}\rightarrow\gamma\tilde{G}$
with a macroscopic lifetime for much of the
GMSB parameter space.

Since decay photons from long-lived particles will have a later arrival time
than prompt photons produced from standard model (SM) sources,
a suitable separation variable is\cite{key-explanation}:
\begin{equation}
\Delta s\equiv(t_{f}-t_{i})-\frac{|\vec{x}_{f}-\vec{x}_{i}|}{c}\label{eq:DeltaseqDeltatxfx0}
\end{equation}
Prompt (SM) photons will produce $\Delta s\equiv 0$ and photons from
long-lived particles $\Delta s > 0$, for perfect measurements.
In general, only neutralinos with both a long lifetime and a low momentum produce large $\Delta s$ photons. Therefore, the
efficiency of a $\Delta s$ cut depends slightly on its momentum distribution, i.e. the production mechanism of the event.
As an example, if a neutralino has an event lifetime of 10~ns, then it has roughly 1\% 
probability to decay in the detector. However if it does decay, the decay photon would pass
a cut of $\Delta s > 3$~ns in 100\% of the cases.
For the CDF detector the system resolution is $\sigma_{\Delta s} \sim1.0$~ns\cite{key-26}.



We investigate separately a $\gamma\gamma$~+~$E_{T}\!\!\!\!\!\!\!/\,\,\,$ 
and a $\gamma$~+~$E_{T}\!\!\!\!\!\!\!/\,\,\,$~+~jets analysis 
for the following reasons: 1) with neutralino lifetimes longer than a  nanosecond,
one or both of the neutralinos can leave the detector
before they decay, 2) gravitinos or the neutralino leaving
the detector provide missing transverse energy, $E_{T}\!\!\!\!\!\!\!/\,\,\,$, 
and 3) in GMSB models the neutralinos are part of cascades from gauginos 
which produce additional particles which, in general, could be identified as jets.
We use PGS\cite{key-6} as a simple detector simulation 
and {\small ISAJET}\cite{key-56} to generate the SUSY masses.
The sensitivity is estimated using the expected 95\%~C.L. cross section upper limits for 2~fb$^{-1}$,
as that is a conservative estimate for the integrated luminosity at the end of run~II.




A $\gamma\gamma$ + $E_{T}\!\!\!\!\!\!\!/\,\,\,$ analysis has
the best sensitivity for low neutralino lifetimes. 
The background for this analysis consists of QCD events with fake $E_{T}\!\!\!\!\!\!\!/\,\,\,$\cite{key-10}.
We find that adding the $\Delta s$ values, $\Delta s_{12}=\Delta s^{\gamma_{1}}+\Delta s^{\gamma_{2}}$,
and selecting signal events with \emph{either} large $E_{T}\!\!\!\!\!\!\!/\,\,\,$ \emph{or}
large $\Delta s_{12}$, either of which is not SM-like,
maximizes the separation of signal and background. 
We find that both the $\Delta s_{12}$
and $E_{T}\!\!\!\!\!\!\!/\,\,\,$ cuts are stable at around 7~ns and 50~GeV for non-zero lifetimes. 

Analogously we proceed with the $\gamma$~+~$E_{T}\!\!\!\!\!\!\!/\,\,\,$~+~jets analysis
which is most sensitive to neutralinos with long lifetime.
The backgrounds are dominated by QCD and $W$+jets\cite{key-8}.
We find that the optimal final selection requirements accept 
events with either large $E_{T}\!\!\!\!\!\!\!/\,\,\,$ \emph{or} large $\Delta s$.
We find a $\Delta s$ cut around 4~ns
which is stable for all masses and lifetimes and $E_{T}\!\!\!\!\!\!\!/\,\,\, > 25$~GeV.

%

\begin{figure}
\centerline{
\psfig{file=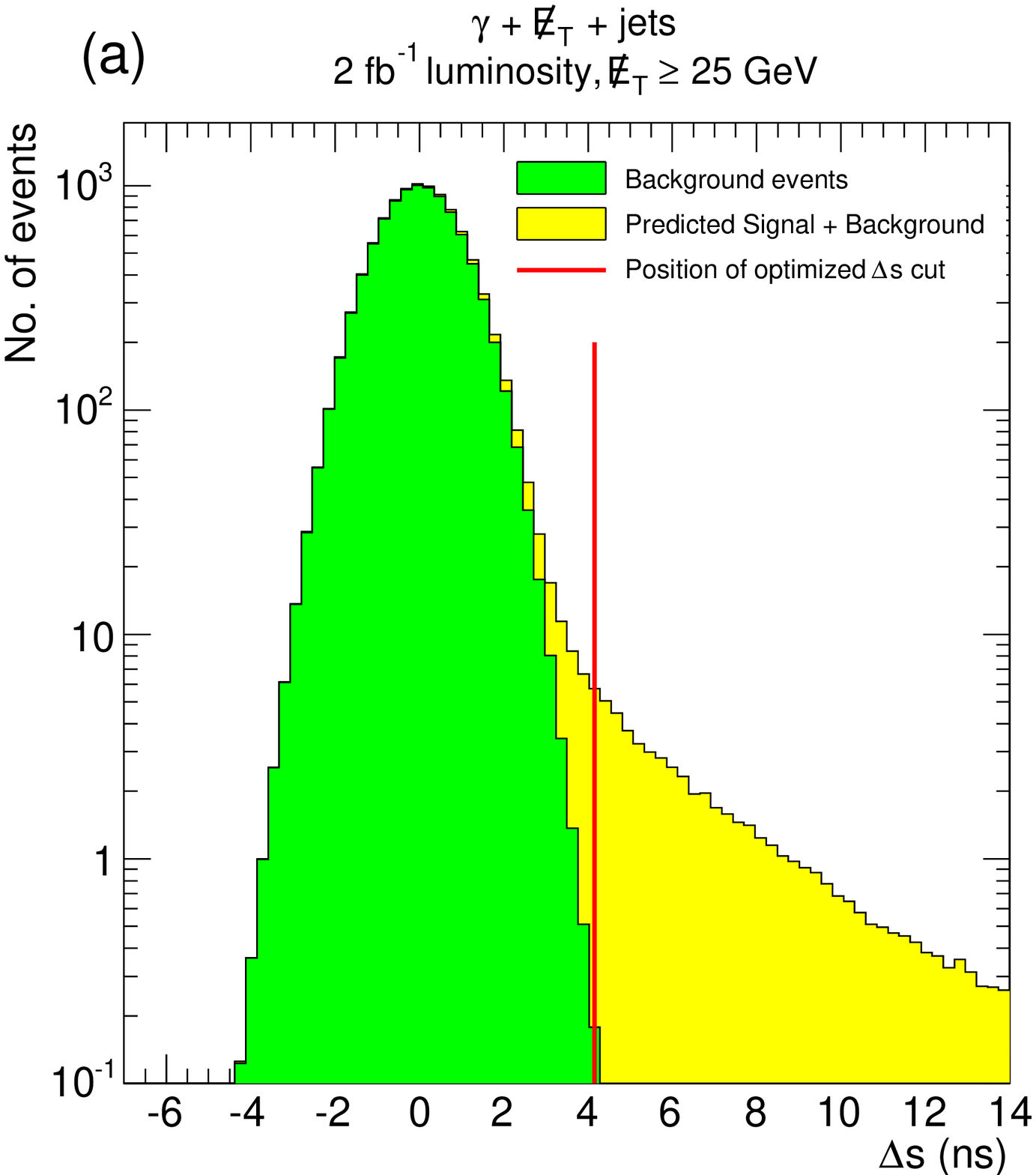,width=4.3cm,height=4.9cm}
\psfig{file=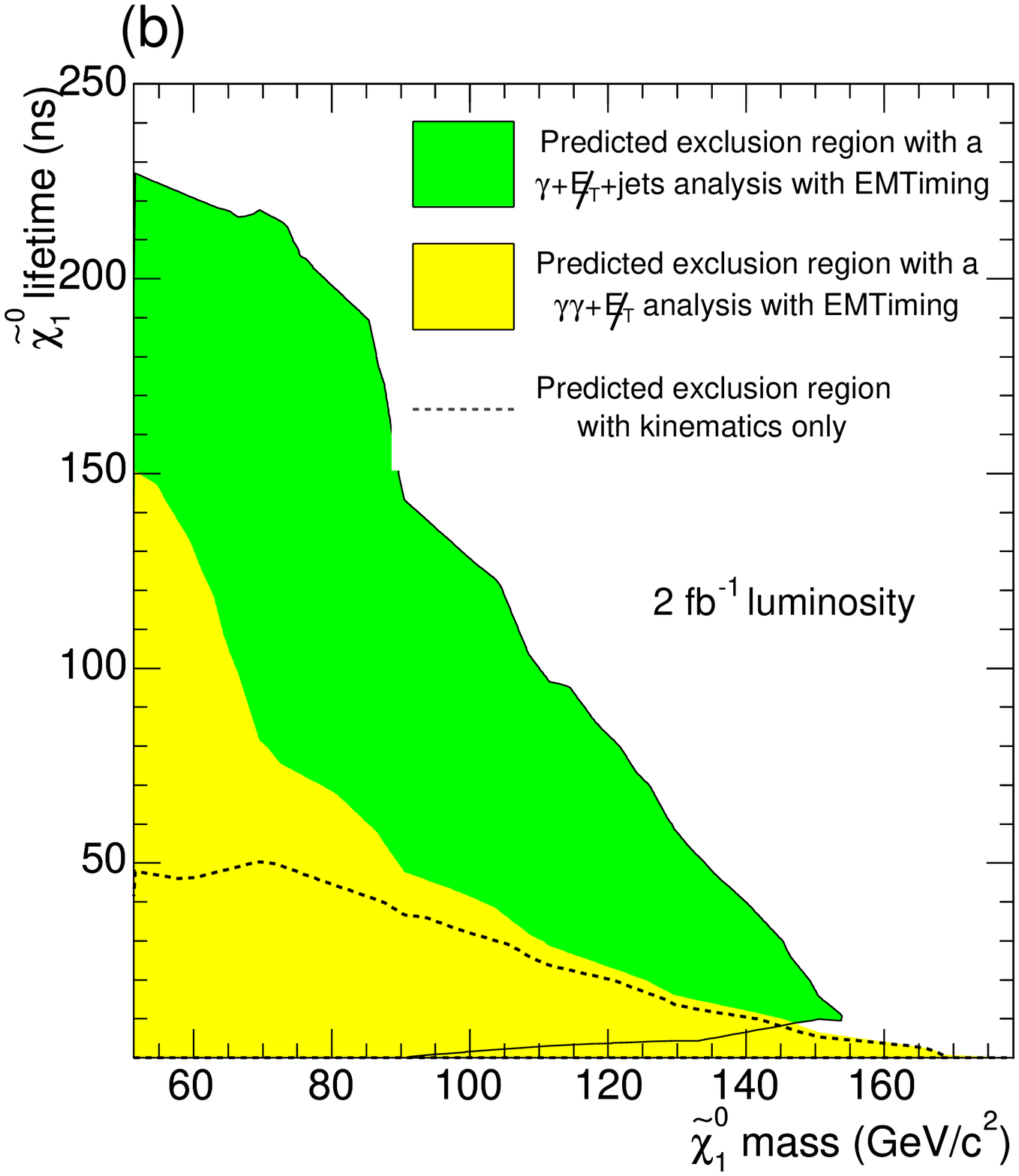,width=4.5cm}
\psfig{file=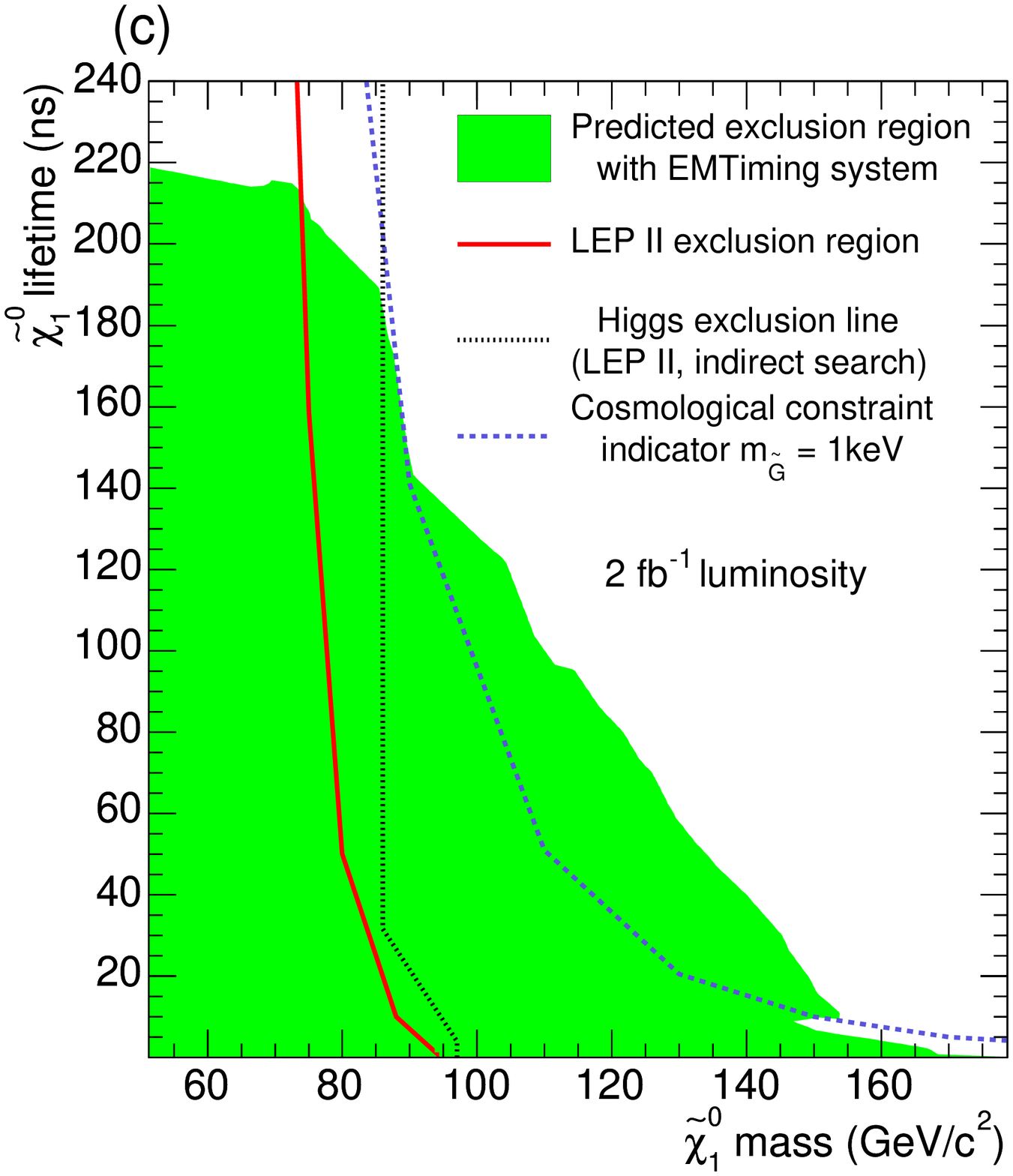,width=4.5cm}
}
\vspace*{8pt}
\caption{\label{cap:xsecmin-gmsb-2000lumi}
Plot (a) shows an example of the distribution of signal (bright) and background (dark) vs. $\Delta s$ 
in the $\gamma$~+~$E_{T}\!\!\!\!\!\!\!/\,\,\,$~+~jets analysis after a baseline $E_{T}\!\!\!\!\!\!\!/\,\,\,$ cut 
of 25~GeV. The line of the optimized $\Delta s$ cut shows that there is good separation between signal and background.
Figures (b) and (c) show the expected 95\%~C.L. exclusion regions as a function of neutralino lifetime
and mass for full GMSB production at $2\ \mathrm{fb}^{-1}$
luminosity. Plot (b) shows separately the exclusion regions for the $\gamma\gamma$~+~$E_{T}\!\!\!\!\!\!\!/\,\,\,$ (bright)
and the $\gamma$~+~$E_{T}\!\!\!\!\!\!\!/\,\,\,$~+~jets analysis (dark). Plot (c) shows the final expected exclusion region 
from the overlap of the two analyses and compares the results to the direct and indirect search limits from ALEPH at 
LEP~II. The $m_{\tilde{G}}=1$~$\mathrm{keV}/c^2$ line is shown as an indicator for the 
theoretically favored region from cosmological considerations.}
\end{figure}

A comparison of the expected cross section limits with the
production cross sections in the GMSB model
gives the neutralino mass vs. lifetime exclusion regions shown in Fig.~\ref{cap:xsecmin-gmsb-2000lumi}
at a luminosity of $2\ \mathrm{fb}^{-1}$.
Timing has the biggest effect at low masses and high lifetimes.
We have also indicated the exclusion regions from ALEPH at LEP~II from both direct
and indirect searches\cite{key-3}. For $2\ \mathrm{fb}^{-1}$ the Tevatron should
significantly extend the sensitivity towards large mass and lifetimes.
The mass exclusion limit at 168~GeV for $\tau_{\tilde{\chi}}$~=~0~ns is comparable to the limit presented
in the D\O~study of displaced photons\cite{key-36}.
Since in most cosmological scenarios the relic density of the gravitino will
overclose the universe if it has a mass of $\ge \mathrm{few}$~$\mathrm{keV}/c^2$~\cite{key-27},
we show the 1~$\mathrm{keV}/c^{2}$ line as an indicator for this theoretically preferred region.
While variations from the chosen GMSB model line have not been further examined, the highest gravitino mass
we can exclude is $\sim$1.7~$\mathrm{keV}/c^{2}$ at $m_{\tilde{\chi}}\approx$~130~\munit\  and $\tau_{\tilde{\chi}}\approx$~60~ns.

We have studied the prospects of using timing information to directly
search for neutral, long-lived particles which
decay to photons, as one can find in SUSY models. We find
that a combination of timing and kinematic requirements provide excellent
rejection against SM backgrounds in complementary fashion. While
the region where timing produces the most additional rejection is
already excluded by ALEPH at LEP~II, the additional handle it provides should allow
the Tevatron in run~II to produce the world's most stringent limits
at masses above 80~\munit\ at high lifetimes. These exclusions have the potential
to come close to cosmological constraints for GMSB models.
The presented prospective results will be tested with the EMTiming system at CDF
whose installation is currently being finished during the CDF shutdown in Fall 2004.

\begin{thebibliography}{0}
\bibitem{key-explanation}In formula (1) $t_{f}-t_{i}$ is the time between the collision and the arrival time of the photon, and $|\vec{x}_{f}-\vec{x}_{i}|$ 
is the distance between the final position of the photon and the collision point.
\end{thebibliography}
\end{document}